\documentclass{article}

\usepackage{arxiv}
\usepackage{graphicx}

\usepackage[utf8]{inputenc} 
\usepackage[T1]{fontenc}    
\usepackage{hyperref}       
\usepackage{url}            
\usepackage{booktabs}       
\usepackage{amsfonts}       
\usepackage{nicefrac}       
\usepackage{microtype}      
\newtheorem{thm}{Theorem}[section]
\newtheorem{proof}{proof}[section]

\title{Estimating Parameters in Mathematical Model for Societal Booms through Bayesian Inference Approach}

\author{
  Yasushi Ota  \\
  Department of Management \\
  Okayama University of Science \\
  Okayama Japan\\ 
  \texttt{yota@mgt.ous.ac.jp}\\
   \And
 Naoki Mizutani \\
  Department of Management \\
  Okayama University of Science \\
  Okayama Japan\\ 
  \texttt{mizutani@mgt.ous.ac.jp} \\
}

\begin{document}
\maketitle

\begin{abstract}
In this study, 
based on our previous study, we examined the mathematical properties, especially the stability of the equilibrium for our proposed mathematical model. 
By means of the results of the stability in this study, 
we also used actual data representing transient booms and resurgent booms, and conducted parameter estimation for our proposed model using Bayesian inference. 
In addition, we conducted a model fitting to five actual data. 
By this study, we reconfirmed that we can express the resurgences or minute vibrations of actual data by means of our proposed model.
\end{abstract}


\section{Introduction}
Booms emerge in many fields 
and are closely tied to our everyday life.
For example, a fashion in clothing, makeup, sports, a movie and food(we call ``societal booms'').
These examples show that ``interesting information'' 
about the individual boom passed at a rapid rate to a large number of people in a short period of time.
Furthermore, in a sense, 
we can regard an infectious disease as a boom, 
like influenza or SARS(we call ``epidemiological booms''), 
which is infected by viruses that are transmitted from person to person.
In the above examples, 
it is the most important things that both ``interesting information'' 
and ``viruses'' are transmitted by some form of contact.
Hence, we considered that a spread with an interest in products, movie, food, etc resemble the transmission dynamics of viruses in ways.

Studies on epidemiological booms that employed mathematical models 
were founded on differential equations in the late 19th century, 
and mathematical modeling using differential equations 
was developed by Kermack and McKendrick in the 1920s. 
This field of study gained attention among researchers 
owing to the spread of emerging infectious diseases, 
such as AIDS in the 1980s, that posed risks in developed countries. 
This research continues to progress (\cite{OdeModel2}, \cite{OdeModel1}, \cite{SIR}). 
On the other hand, there are many studies that researched societal booms from a sociological or psychological perspective, but few were conducted from a mathematical point of view. However, in recent years, companies have concentrated their marketing efforts on the development of hit products that emphasize customer taste and trend analysis by using social networking services (SNS) such as Twitter and Facebook. 

In this study, 
we focus on societal booms and we develop research for it.
Ishi et al. \cite{ishii} have derived a mathematical model for the ``hit" phenomenon in entertainment within a society, 
which is presented as a stochastic process of human dynamics interactions.
Ishi et al. have performed calculations using their proposed equation for many movies in the Japanese market.
Moreover, Nakagiri and Kurita \cite{nakagiri} conducted one study that focused on societal booms. 
Nakagiri et al. used a system of simultaneous linear differential equations to develop a mathematical model to describe problems in societal booms, and performed a model fitting to actual data. 
The mathematical model proposed in this study is simple but extremely versatile. 
Additionally, Ueda and Asahi \cite{ueda} expanded on the model developed by Nakagiri et al. to conduct an analysis using actual data by constructing and verifying a model of the changing interests among Twitter users. 
In \cite{Yasu}, 
we proposed the mathematical 
boom model developed by Nakagiri et al. in consideration of the SIR model \cite{SIR}, 
which is a leading idea to describe biological booms 
such as viral infections, 
and the Diffusion of Innovation theory \cite{Rogers} 
proposed by sociologist E.M. Rogers.
In this study, we examine the stability of the equilibrium of our proposed model in \cite{Yasu}.
Moreover, using actual booms data we evaluate the parameters and examine the fit of our proposed model.

This study is divided into six parts.
In Section 2, we explain the ideas at the core of our proposed mathematical model for Societal Booms, which was proposed in our previous study.
In Section 3, 
we investigate the stability of the equilibrium point to our proposed model and derive the sufficient condition for parameters of our proposed model.
In Section 4, we explain the Bayesian inference method, which was used to estimate the parameters of our proposed model, and discusses the numerical exploration of the posterior state space by the MCMC method. Moreover, we introduce the coefficient of determination that forms the standard for the fit.
In Section 5, we evaluate the parameters of our proposed model and examine fitting our proposed model to actual data, using five actual data for societal booms.

\section{Mathematical Model}
In this section, 
we explain a mathematical model for societal booms which was derived in \cite{Yasu}.

\subsection{Three key points to derive our proposed model} 
Here, we explain the three key points 
which were discussed to derive our mathematical model.

The first point is the contact. 
Infectious disease epidemics such as influenza are thought 
to occur when a virus invades and infects a healthy person's body 
from contact with an infected person. 
In our proposed model, 
we define ``interesting information'' to be a ``virus'' , 
which is transmitted from people in an on-boom state to those whom 
the boom has not reached yet (pre-boom).
Thus, our proposed model incorporates 
the perspective of the contact, 
which was not considered in \cite{nakagiri}.

The second point is the time delay. 
The Diffusion of Innovation theory, developed by E.M. Rogers, 
separates consumers into five categories based on the speed at which people are likely to adopt innovation (innovators, early adopters, early majority, late majority, and laggards). 
Based on this theory, 
we think that time lags exist in the adoption of booms by people in a social system, and thus developed a model that considers the effects of a time delay.
Hutchinson \cite{Hut} suggested the following logistic equation with time delay $\tau$. 
This equation shows that the solution does not fluctuate 
monotonously but exhibits complex behavior such as oscillatory behavior depending on the magnitude of the time delay:
\begin{equation}
\displaystyle \frac{dx(t)}{dt} = \alpha x(t) \left(1-\frac{x(t-\tau)}{K}\right) \hspace{5ex} (\alpha, K>0)
\label{Hut}
\end{equation}

Furthermore, the biologist R.M. May \cite{May} regarded (\ref{Hut}) as a mathematical model 
that expresses 
the temporal changes of the herbivorous animal population $x(t)$, 
and asserted that the biological definition of 
a time delay was ``the time required 
for the regeneration of plants that is suitable for animals to eat''. 
Additionally, 
May received acclaim for fitting results 
from an experiment on the Australian sheep blowfly (Lucilia cuprina) conducted by Nicholson \cite{Nich}. 
Based on these experimental results,  
we regarded the definition of a time delay for societal booms 
as ``the time required for a boom adopter associated with contact 
and resurgence to pick up a boom and take action'', 
and incorporated the concept of the time delay 
into the derivation of a mathematical model.  

The last point is the existence of influencer and ``Sakura\footnote{``Sakura'' are people who were compelled to boom state.}''. 
In the current landscape, 
companies actively employ influencer and ``Sakura'' 
in their marketing strategies. 
Our proposed model expresses their presence by depicting a resurgence caused by opinion leaders and forced changes in the number of people. 

As described above, 
our proposed model is a natural extension of the boom model 
developed in \cite{nakagiri} and is derived 
from the above three perspectives. 
We expect that the model will be able to 
capture various types of boom data.

\subsection{Mathematical model for societal booms}
In \cite{Yasu}, we proposed a new mathematical model to explain societal booms based on the background described above.  
Here, according to \cite{Yasu} we derive the mathematical model for societal booms.

First, we assume the state of the boom participants at any given time to be one of the following:
\begin{itemize}
  \item{\textbf{State 1 pre-boom}: Condition where there is a potential to adopt a boom}
  \item{\textbf{State 2 on-boom}: Condition where the boom is captured}
  \item{\textbf{State 3 rooted boom}: Condition where the boom is retained}
  \item{\textbf{State 4 unrooted boom}: Condition where the boom did not take off}
\end{itemize}

Furthermore, at a given time $t$, we assume 
the number of boom participants in each state to be $y_1(t), y_2(t), y_3(t)$, and $y_4(t)$ respectively, for States $1-4$. 
Then we represent the changes of a customer's state by using the following equations:
\begin{eqnarray}
\left\{ \begin{array}{l}
\displaystyle \frac{dy_1(t)}{dt} = -\alpha y_1(t)y_2(t-\tau_1)-\delta y_1(t)+\varepsilon y_2(t-\tau_2)-\zeta \\
\mbox{} \\
\displaystyle \frac{dy_2(t)}{dt} = \alpha y_1(t)y_2(t-\tau_1)-(\beta+\gamma)y_2(t)+\delta y_1(t) \\
\mbox{} \\
\displaystyle \frac{dy_3(t)}{dt} = \beta y_2(t)-\varepsilon y_2(t-\tau_2)+\zeta \\
\mbox{} \\
\displaystyle \frac{dy_4(t)}{dt} = \gamma y_2(t) 
\end{array} \right.
\label{maineq}
\end{eqnarray}
Here, the variables $\alpha, \beta, \gamma, \delta, \varepsilon, \zeta$ and 
$\tau_1, \tau_2$ in (\ref{maineq}) respectively 
represent the rate of transmission (``infection'') of 
the boom among people in a pre-boom state per unit time, 
rate of retention among people in an on-boom state, 
rate of people who quit the boom, 
adoption rate of the boom by people 
in a pre-boom state, 
rate of resurgence from rooted boom to pre-rooted state, 
and the production rate of people in an rooted boom state. 
$\tau_1$ and $\tau_2$ are parameters that show the time delay, 
and $\tau_1 < \tau_2$. 
Here, $\alpha>0, \beta + \gamma > 0, \delta>0$. 
In particular, $\alpha y_2(t-\tau_1)$ is called 
as infectivity and is an important indicator 
that characterizes the boom model. 

In this model, 
the rate at which people go from a pre-boom state to an on-boom state 
is proportional to 
the number of people in a pre-boom state 
and the number of people who changed to an on-boom state before $\tau_1$ 
(the first term of the first equation in (\ref{maineq})), 
and we assume that people in a pre-boom state naturally adopt booms 
at a fixed rate (the second term of the first equation in (\ref{maineq})). Moreover, it expresses the resurgence of the boom 
by transferring the rate of people 
who became on-boom prior to $\tau_2$
(the third term of the first equation in (\ref{maineq})). 
In addition, $\zeta$ expresses the ratio of people in an rooted boom state who were compelled to enter that state (in other words, ``Sakura'').
Furthermore, the given ratio of people in an on-boom state enter an rooted or unrooted state(the second and third term of the second equation in (\ref{maineq})). 

\section{Stability of the equilibrium point }

In this section, we analyze the stability of equilibrium by 
investigating location of the roots of associated characteristic equation.
Here, since the right side of each equation in system (\ref{maineq}) 
consists of only $y_1, y_2$, we study the stability of the equilibrium point for the first equation and the second equation in system (\ref{maineq}).
The first equation and the second equation in system (\ref{maineq}) possess a trivial equilibrium $E_0=(0,0)$, if $\zeta=0$, and non trivial equilibrium
\begin{eqnarray*}
E_1 = (y^*_1,y^*_2)=\left(\frac{(\beta+\gamma)\zeta}{\alpha\zeta+\delta(\beta+\gamma-\varepsilon)}, \ \frac{\zeta}{\beta+\gamma-\varepsilon}\right),
\end{eqnarray*}
if $\zeta \neq 0$.
In this study, we analyze the stability of non trivial equilibrium $E_1$.

First, upon the following change of functions and substitutions
\begin{eqnarray*}
\left\{ \begin{array}{l}
a(t)=y_1(t)-y^*_1, \\
b(t)=y_2(t)-y^*_2,
\end{array}  \right.
\end{eqnarray*}
system (\ref{maineq}) becomes
\begin{eqnarray}
\left\{
\begin{array}{l}
\displaystyle \frac{da}{dt}=-\alpha a(t)b(t-\tau_1)-\alpha a(t)y^*_2-\alpha y^*_1b(t-\tau_1)-\alpha y^*_1y^*_2 \\
\hspace{35ex} -\delta a(t)-\delta y^*_1+\varepsilon b(t-\tau_2)+\varepsilon y^*_2+\zeta, \\
\mbox{}  \\
\displaystyle \frac{db}{dt}=\alpha a(t)b(t-\tau_1)+\alpha a(t)y^*_2+\alpha b(t-\tau_1)y^*_1+\alpha y^*_1y^*_2 \\
\hspace{35ex} +\delta a(t)+\delta y^*_1-(\beta+\gamma)b(t)-(\beta+\gamma)y^*_2.
\end{array}
\right.
\label{maineq2}
\end{eqnarray}

\noindent Since $y_1^*, y_2^*$ are the equilibrium point, (\ref{maineq2}) becomes the following form:
\begin{eqnarray*}
\left\{ \begin{array}{l}
\displaystyle \frac{da}{dt}=-(\alpha y^*_2+\delta)a(t)-\alpha y^*_1b(t-\tau_1)+\varepsilon b(t-\tau_2), \\
\mbox{} \\
\displaystyle \frac{db}{dt}=(\alpha y^*_2+\delta)a(t)+\alpha y^*_1b(t-\tau_1)-(\beta+\gamma)b(t).
\end{array}  \right.
\end{eqnarray*}

\noindent Regarding this equation as system ODE with the equilibrium point $(0,0)$, 
we have 
\begin{eqnarray}
  \frac{d}{dt} \left(
    \begin{array}{c}
      a(t) \\
      b(t) \\
    \end{array}
  \right)
  = A_1
  \left(
    \begin{array}{c}
      a(t)  \\
      b(t)  \\
    \end{array}
  \right)  
  + A_2
    \left(
    \begin{array}{c}
      a(t-\tau_1)  \\
      b(t-\tau_1)  \\
    \end{array}
  \right) 
  + A_3
     \left(
    \begin{array}{c}
      a(t-\tau_2)  \\
      b(t-\tau_2)  \\
    \end{array}
  \right),
  \label{ODE_0}
\end{eqnarray}
where 
\begin{eqnarray*}
A_1  =\left(
    \begin{array}{cc}
      -\alpha y^*_2-\delta   & 0   \\
      \alpha y^*_2+\delta  &  -(\beta+\gamma)  \\
    \end{array}
  \right), \
A_2  =\left(
    \begin{array}{cc}
      0   & -\alpha y^*_1   \\
      0  &  \alpha y^*_1  \\
    \end{array}
  \right), \
A_3  =\left(
    \begin{array}{cc}
      0   & \varepsilon   \\
      0  &  0  \\
    \end{array}
  \right).
\end{eqnarray*}
Therefore, the characteristics equation of the above system ODE takes the following form:
\begin{eqnarray*}
\mbox{det}(\lambda I - A_1-A_2 e^{-\tau_1 \lambda}- A_3 e^{-\tau_2 \lambda}) = 0,
\end{eqnarray*}
That is, 
\begin{eqnarray*}
\left|\left(
  \begin{array}{cc}
     \lambda+(\alpha y^*_2+\delta)  & \alpha y^*_1e^{-\tau_1\lambda}-\varepsilon e^{-\tau_2\lambda}   \\
     -(\alpha y^*_2+\delta)  & \lambda+(\beta+\gamma)-\alpha y^*_1e^{-\tau_1\lambda}   \\
  \end{array}
\right)\right| = 0, \\
\end{eqnarray*}
which implies
\begin{equation}
\lambda^2 + (\beta+\gamma+\delta - \alpha y_1^* e^{-\tau_1 \lambda}
+ \alpha y_2^*)\lambda +(\alpha y_2^* +\delta)(\beta +\gamma-\varepsilon
e^{-\tau_2 \lambda}) = 0.
\label{ch-eq-th}
\end{equation}

\begin{thm}
\label{thm}
We consider Esq. (\ref{maineq}), where $\alpha, \beta, \gamma, \delta$, and $\tau_1, \tau_2$ are non-negative numbers and $\varepsilon, \zeta$ are real numbers. 
And let us assume that 
\begin{equation}
\beta+\gamma - \varepsilon > 0 , 
\label{condi1}
\end{equation}
\begin{equation}
\frac{\alpha\zeta+\delta(\beta+\gamma-\varepsilon)}{\alpha\zeta(\beta+\gamma)}>\frac{\tau_1}{2},
\label{condi3}
\end{equation}
moreover,
\begin{equation}
\alpha\zeta + \delta (\beta+\gamma-\varepsilon)>0, \\
\label{condi2}
\end{equation}
\begin{equation}
\frac{1}{\varepsilon}\left(\frac{\delta(\beta+\gamma)(\beta+\gamma-\varepsilon)^2}{(\alpha\zeta+\delta (\beta+\gamma-\varepsilon))^2}+1\right)>\tau_2, 
\label{condi4}
\end{equation}
\mbox{or}
\begin{equation}
\alpha\zeta + \delta (\beta+\gamma-\varepsilon)<0, \\
\label{condi6}
\end{equation}
\begin{equation}
\frac{1}{\varepsilon}\left(\frac{\delta(\beta+\gamma)(\beta+\gamma-\varepsilon)^2}{(\alpha\zeta+\delta (\beta+\gamma-\varepsilon))^2}+1\right)<\tau_2. 
\label{condi8}
\end{equation}

\noindent Then we shall show that every root of the characteristic equation (\ref{ch-eq-th}) must have negative real part.
\end{thm}

\pagebreak

\begin{proof}
Suppose for contradiction that 
$\lambda=\mu+i\omega$
is a root of (\ref{ch-eq-th}) with $\mu \ge 0, \omega\neq 0.$
Then, 
\begin{eqnarray*}
\mbox{Im}[\{\lambda^2+(\beta+\gamma+\delta+\alpha(y^*_2-y^*_1e^{-\tau_1\lambda})\lambda+(\alpha y^*_2+\delta)(\beta+\gamma-\varepsilon e^{-\tau_2\lambda})\}/\omega] \\
\mbox{} \\
=\mbox{Im}[\{(\mu+i\omega)^2+(\beta+\gamma+\delta+\alpha y^*_2-\alpha y^*_1e^{-\tau_1(\mu+i\omega)}(\mu+i\omega) \hspace{14.5ex}\\
+(\alpha y^*_2+\delta)(\beta+\gamma+\varepsilon e^{-\tau_2(\mu+i\omega))}\}/\omega] \\
\mbox{} \\
=\{2\mu\omega+(\beta+\gamma+\delta+\alpha y^*_2-\alpha y^*_1e^{-\tau_1\mu}\cos\tau_1\omega)\omega \hspace{25ex}\\
+\alpha y^*_1e^{-\tau_1\mu}\sin\tau_1\omega\mu+(\alpha y^*_2+\delta)\varepsilon e^{-\tau_2\mu}\sin\tau_2\omega\}/\omega 
\mbox{} \\
=2\mu+\beta+\gamma+\delta+\alpha y^*_2-\alpha y^*_1e^{-\tau_1\mu}\cos\tau_1\omega \hspace{30.5ex}\\
+\alpha y^*_1e^{-\tau_1\mu}\frac{\sin\tau_1\omega}{\omega}\tau_1\mu+(\alpha y^*_2+\delta)\varepsilon e^{-\tau_2\mu}\frac{\sin\tau_2\omega}{\omega}\tau_2. 
\end{eqnarray*}

\noindent From $\tau_i>0, \ \mu>0$,
\begin{eqnarray*}
e^{-\tau_i\mu}<1, \hspace{2ex} -1\leq \frac{\sin\tau_i\omega}{\tau_i\omega}\leq 1 \hspace{2ex}(i=1,2),
\end{eqnarray*}
therefore, we have
\begin{eqnarray*}
f(\mu)>2\mu+\beta+\gamma+\delta+\alpha y^*_2-\alpha y^*_1-\alpha y^*_1\tau_1\mu-(\alpha y^*_2+\delta)\varepsilon\tau_2 \\
\mbox{} \\
=(2-\alpha y^*_1 \tau_1) \mu + \beta + \gamma + \delta 
+ \alpha(y^*_2-y^*_1)-(\alpha y^*_2 + \delta) \varepsilon \tau_2.
\end{eqnarray*}

At first, from $\mu >0$ and (\ref{condi3}), we have
\begin{eqnarray*}
2-\alpha y^*_1 \tau_1 > 
2 - \alpha \displaystyle \frac{\zeta(\beta+\gamma)}{B} \frac{2B}{\alpha \zeta(\beta+\gamma)}=0.
\end{eqnarray*}

\noindent Next,
\begin{eqnarray*}
\beta+\gamma+\delta+\alpha(y^*_2-y^*_1)-(\alpha y^*_2+\delta)\varepsilon \tau_2 \hspace{35ex}\\
\mbox{} \\
=\displaystyle\frac{(\beta+\gamma+\delta)AB+\alpha(\zeta B - (\beta+\gamma)\zeta A)-(\alpha\zeta+\delta A)\varepsilon B \tau_2}{AB} \hspace{10ex} \\
\mbox{} \\
=\frac{\{(\beta+\gamma+\delta)B - \alpha\zeta(\beta+\gamma) - \delta \varepsilon B \tau_2 \}A + (\alpha \zeta - \alpha \zeta \varepsilon \tau_2)B}{AB} \hspace{8.5ex}\\
\mbox{} \\
=\frac{\{\delta B(1-\varepsilon\tau_2) +(\beta+\gamma)\delta A \}A + \alpha \zeta B (1-\varepsilon\tau_2)}{AB} \hspace{19.5ex}\\
\mbox{} \\
=\frac{(\beta+\gamma)\delta A^2 + (1-\varepsilon\tau_2)B^2}{AB}. \hspace{38ex}
\end{eqnarray*}
From (\ref{condi2}), (\ref{condi4}) or (\ref{condi6}), (\ref{condi8}), 
we have
\begin{eqnarray*}
\frac{(\beta+\gamma)\delta A^2 + (1-\varepsilon\tau_2)B^2}{AB}
>0.
\end{eqnarray*}

This is contradiction shows that every $\lambda$ has negative real part.
Hence every solution of Esq. (\ref{maineq}) tends to zero exponentially as $t \to \infty$.
\end{proof}

\pagebreak

\section{MCMC methods}

In Bayesian inference approach, 
the complicated and intractable probabilistic models 
can be estimated 
by numerical sampling methods such as MCMC, 
which has been widely applied in recent years.
The details of MCMC methods are given 
in Robert and Casella \cite{Robert-Casella}.

Monte Carlo simulation generates pseudo--random numbers 
for exploring posterior distributions.
In the MCMC algorithm, 
the pseudo--random number is a Markov chain. 
The MCMC algorithm exploits 
the property of a Markov chain to generate 
pseudo--random numbers from a posterior distribution,  
even for a complicated model.
It first constructs an ergodic Markov chain with a stationary distribution equaling the target distribution. 
By iterating the Markov chain transitions from suitable initial value, it eventually obtains the target distribution. 

In this paper, we employs a typical MCMC algorithm called the Metropolis--Hastings (M--H) algorithm (see Metropolis et al. \cite{Metropolis}; Hastings \cite{Hastings}). 
The M--H Algorithm given below builds its Markov chain by accepting or rejecting samples extracted from a proposed distribution. 
This algorithm is generally used 
in Bayesian inference approach (cf. \cite{mcmc-springer}).

\vspace{2ex}

\noindent\textbf{\underline{M--H Algorithm}}
\begin{itemize}
  \item{\textbf{Step1}: Generate $\theta'\sim q(\cdot|\theta_k)=N(\theta_k,\sigma_\theta^2)$ (the normal distribution) with a given stander derivation $\sigma>0$ for given $\theta_k$.}
  \item{\textbf{Step2}: Calculate the choice $\alpha(\theta',\theta_k)=\min\left\{1,f(\theta'|Y)/f(\theta_k|Y)\right\}$.}
  \item{\textbf{Step3}: Update $\theta_k$ as $\theta_{k+1}=\theta'$ with probability $\alpha(\theta',\theta_k)$ but otherwise set $\theta_{k+1}=\theta_k$.}
\end{itemize}

\vspace{2ex}

By running the M--H algorithm, 
we can sample the distribution $f(\theta|Y)$, 
and usually the mean value of $\theta_j$, after a given burn-in time $j^*$.

\section{Using Real Data to Evaluate Validity of Proposed Model}
\textbf{Parameter Estimation Steps(PES)} below is the procedure for the parameter estimation for the model fitting to actual data conducted in this study. 
The values referred in Step 2 of PES were determined from actual data as follows:

\vspace{2ex}

\noindent\textbf{\underline{Parameter Estimation Steps(PES)}}
\begin{itemize}
  \item{\textbf{Step1}: Set appropriate initial values for $\alpha \sim \varepsilon$.}
  \item{\textbf{Step2}: Set initial values for $\zeta$\footnotemark[1] and $\tau_1, \tau_2$ from actual data.}
  \item{\textbf{Step3}: Apply algorithm 1(MCMC-MH) using $20,000$ iterations as the standard to determine the values of $\alpha \sim \varepsilon$.}
  \item{\textbf{Step4}: Visually confirm the fitting to actual data, and confirm the coefficients of determination.}
  \item{\textbf{Step5}: Fine-tune the values of $\zeta$ and $\tau_1, \tau_2$.}
  \item{\textbf{Step6}: Repeat steps 3-5 to determine the final parameters based on the sufficient condition in Theorem \ref{thm}, changes in the coefficients of determination and visual observation.}
\end{itemize}
\footnotetext[1]{For ``Reiwa'' data, the initial value of the "Sakura'' parameter $\zeta$ was set to about $5\%$ of the peak figure for each data set, since we assume the presence of people who keep having interest toward May 1 on which Prince Naruhito enthroned the throne.}

\vspace{2ex}

We determined the values for time delay parameters $\tau_1$ and $\tau_2$ from the actual data, referencing events that occurred prior to each cultural phenomenon. 
In particular, for $\tau_1$ we observed the first peak and set the initial value. 
Moreover, for $\tau_2$ we set the initial value to be the duration between this point of change to a large peak, on unusual changes such as peaks created from resurgence.

\subsection{Coefficient of determination}

In this study, we introduce the coefficient of determination $\mbox{R}^2$ as an index, which indicates the percentage change between the solution of the proposed model and an actual data, as follows:
\begin{equation}
R^2 = 1-\frac{\displaystyle\sum_{i=1}^{n} (E_i - \bar{E})^2}{\displaystyle\sum_{i=1}^n (Y_i-\bar{Y})^2}
\label{coe_deter}
\end{equation}

Then, from the following form(\ref{coe_deter}), 
the range is from $0$ to $1$ and the $\mbox{R}^2$ of 1 means an actual data can be predicted without error from the solution of the proposed model.
Here, $\bar{Y}$ is a mean of an actual data, and $E_i=Y_i-F_i^{\theta}$, $\bar{E}$is a mean of $E_i$.

\subsection{Comparisons to Prior Studies}

\begin{figure}[h]
\centering
  \includegraphics[width=80mm]{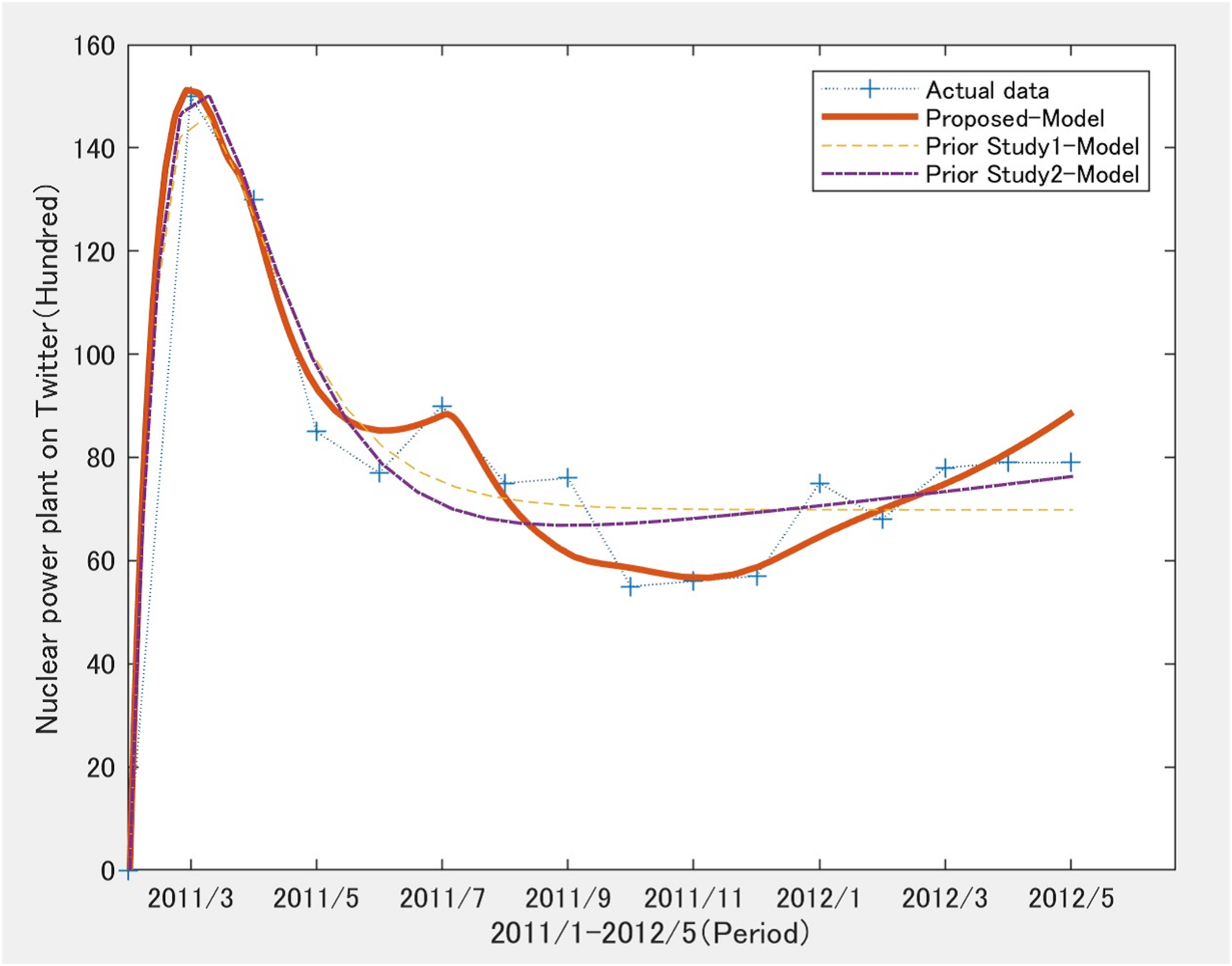}
  \caption{Comparison to prior studies}
  \label{compe}
\end{figure}

In this section, according to \cite{Yasu}, 
we compare the solution of our proposed model to the solutions of the proposed model in \cite{nakagiri}(Prior Study1) and \cite{ueda}(Prior Study2), 
to verify the superiority of our proposed model which is used in this study.
We employed the number of tweets related to a nuclear power plant as the basis for comparison, in Prior Study2 used this data to compare their study with Prior Study2. 
Here, the parameter values were estimated 
according to Algorithm 1 above.
Figure \ref{compe} also shows that the model developed in Prior Study1(dashed line) does not accurately express the resurgence following the decline.
The model developed in Prior Study2(dashed-dotted line) captures the resurgence following the decline but could not reproduce the second peak or the dip in the middle stage.
On the other hand, our proposed model(solid line) is able to capture the resurgence following the decline, the second peak, and the midway dip that were not expressed by the models in Prior Studies1 and 2.
The accuracy of the model is not only visually evident 
but can also be seen from the value of the coefficient of determination. 
The figure is $\mbox{R}^2=0.9609$ for our proposed model compared to $\mbox{R}^2=0.9129$ in Prior Study1 and $\mbox{R}^2=0.9238$ in Prior Study2.
Therefore, it can be said that 
there is high utility in using our proposed model 
to fit the actual data, as we show below.

\pagebreak

\subsection{Model Fitting to Actual Data}

\subsubsection{La La Land}

La La Land is a buzz-worthy American romantic musical film.
La La Land premiered at the 73rd Venice International Film Festival on August 31, 2016, and was released in the United States on December 9, 2016 and has grossed \$446 million to data.
Moreover it received 14 Academy Award nominations at the 89th Academy Awards and won in six categories, including Best Director and Best Actress.
From the above points, we regards the film ``La La Land'' to be appropriate for reflecting a transitory boom and used the data as the subject for testing.

In this study, we used the number of user reviews left on Yahoo!Japan Movies during the period from August 2016 to July 2018, 
including March 2017 when La La Land won in six categories at the 89th Academy with respect to the fit proposed model to actual data. 

\begin{figure}[h]
\centering
  \includegraphics[width=80mm]{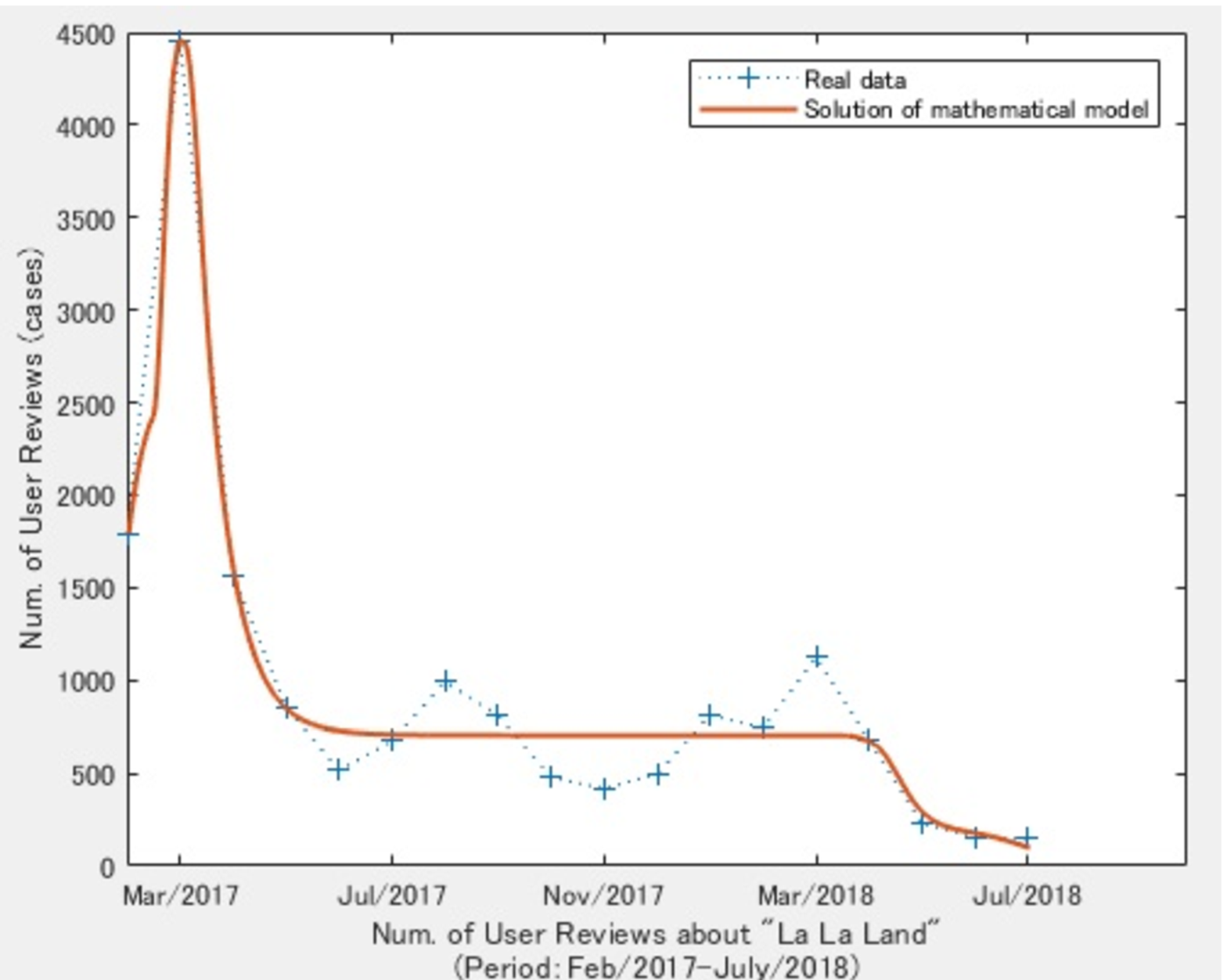}
  \caption{Fitting to La La Land data}
  \label{lalaland}
\end{figure}

The actual data in Figure \ref{lalaland} show a decline after the first peak on March 2017 and finally decrease after two small peaks.
The graph from the proposed model shows a high degree of fit to the first peak. 
In addition,
we observe that it is able to largely replicate the subsequent curves including the leveling-off and decrease rapidly towards the end of the period after two small peaks. 
In addition, the coefficient of determination, which is the measure of 
how well a model explains the data, shows a high value at $\mbox{R}^2 = 0.9688$; 
this reveals that the proposed model has a high degree of fitness. 
On the other hand, 
we observe that fitting to two small peaks in the middle of the period is not good. 

\begin{figure}
\centering
  \includegraphics[width=50mm]{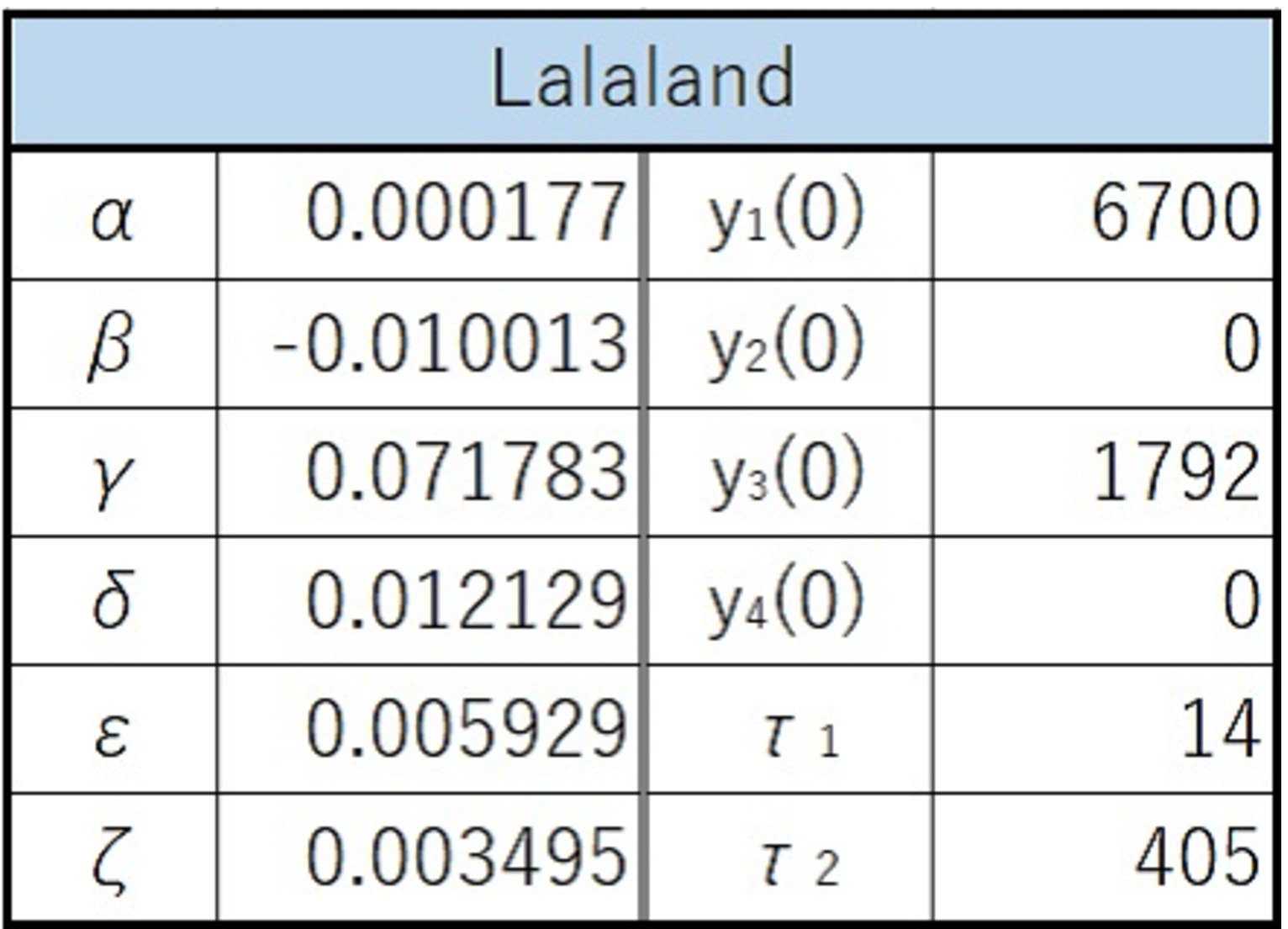}
  \caption{Various parameter values}
  \label{lalaland_para}
\end{figure}

\pagebreak

\subsubsection{``Ten-nensui (Pure Water)''}
Suntory ``Ten-nensui'' is best-selling mineral water 
which made with water from renowned water resources in Japan, 
including the Minami-Alps.
All Suntory Ten-nensui products are made from ``soft water'', clear in color, and beloved for their refreshing taste.

In this study, we used Twitter data from before and after the product launch on 4/17/2019 to test the effectiveness of the ``GREEN TEA CAMPAIGN'' which is the new product of 
``Ten-nensui (Pure water)'' with respect to 
the fit proposed model to actual data. 

\begin{figure}[h]
\centering
  \includegraphics[width=80mm]{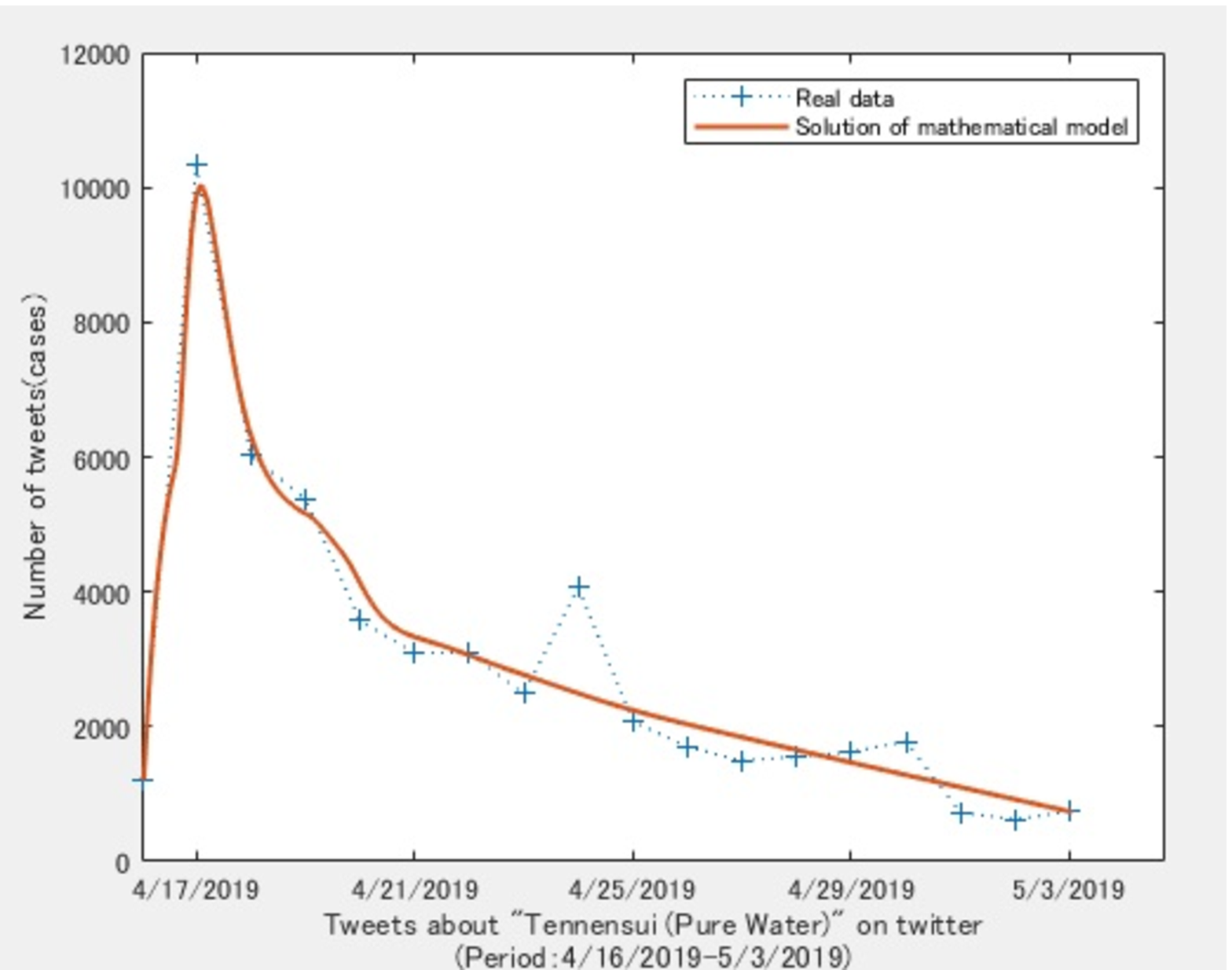}
  \caption{Fitting to ``Tennensui'' Pure-Water data}
  \label{pure_water}
\end{figure}

The actual data in Figure \ref{pure_water} show a decline after the first peak. 
After that, it has a small second peak once more, but it decreases again.
Similar to the other examples, the proposed model shows a high degree of fit to the first peak. 
In addition, we observe that the graph from the proposed model is able to well reconstruct the processes of declining and even expresses 
the small decline that occurs from 4/18/2019 to 4/21/2019. 
The high degree of accuracy in the fitness is evident from the large coefficient of determination, $\mbox{R}^2 = 0.9612$.
On the other hand, we observe that the graph from the proposed model can't express well the second peak in the middle of the period. 

\begin{figure}
\centering
  \includegraphics[width=50mm]{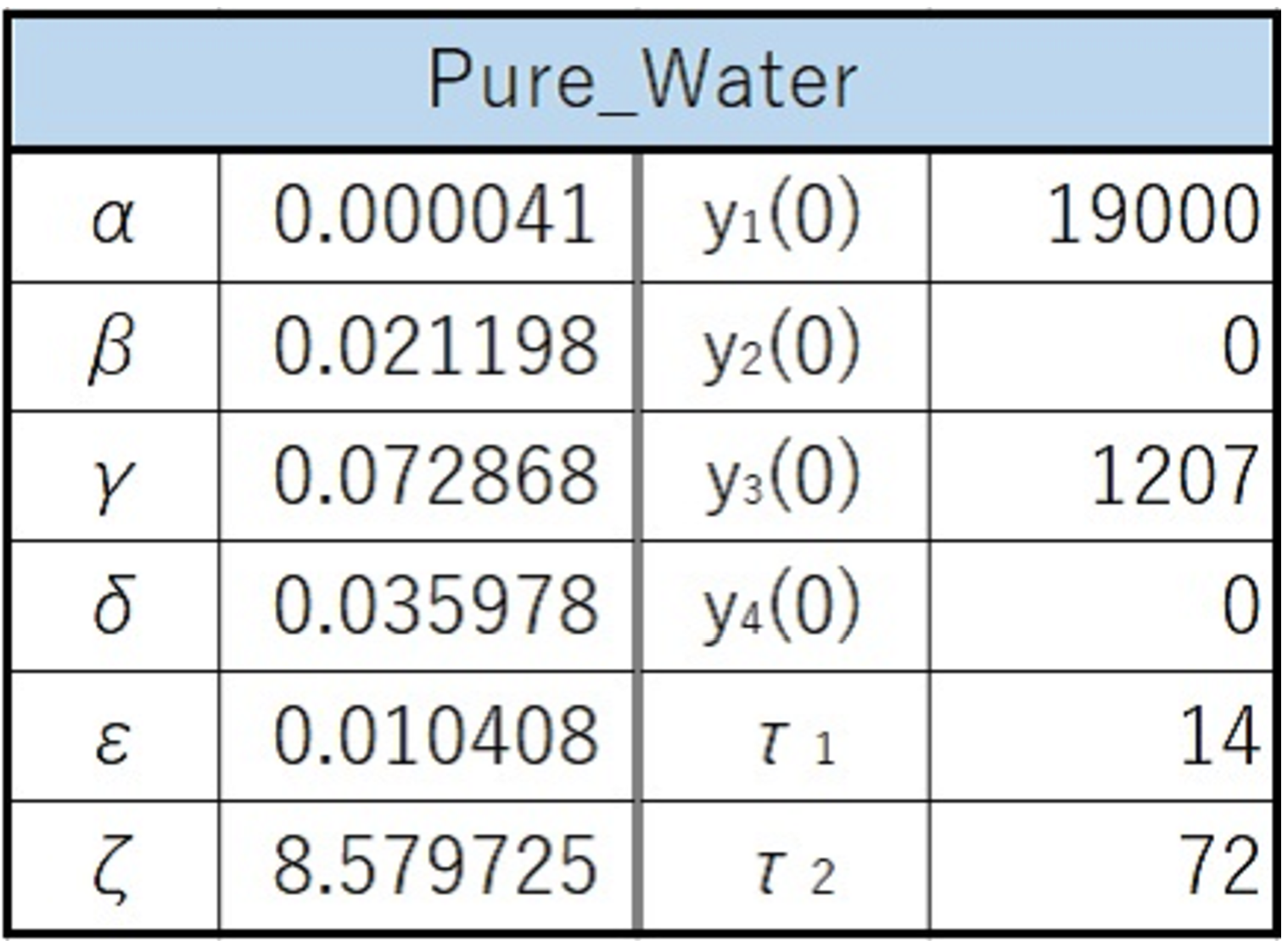}
  \caption{Various parameter values}
  \label{pure_water_para}
\end{figure}

\pagebreak

\subsubsection{Honkirin Beer}
Honkirin Beer is a Happoshu (low-malt) beer that was introduced on March 13, 2018. 
In a cost-conscious environment, 
Honkirin Beer became the biggest hit among new beer releases in FY 2018 owing to its high quality and low price.  
According to the brewer Kirin, 
Honkirin Beer underwent a product renewal 
in mid-January 2019 for an even more refined authentic taste. 
As a result, the product logged a record sales volume in February (1.12 million cases), second only to its release in March 2018 (1.17 million cases). 

Here, too, we used Twitter data 
from before and after the product launch on 4/22/2019 to test the effectiveness of the Honkirin Beer campaign 
with respect to the fit of the proposed model to actual data. 

\begin{figure}[h]
\centering
  \includegraphics[width=80mm]{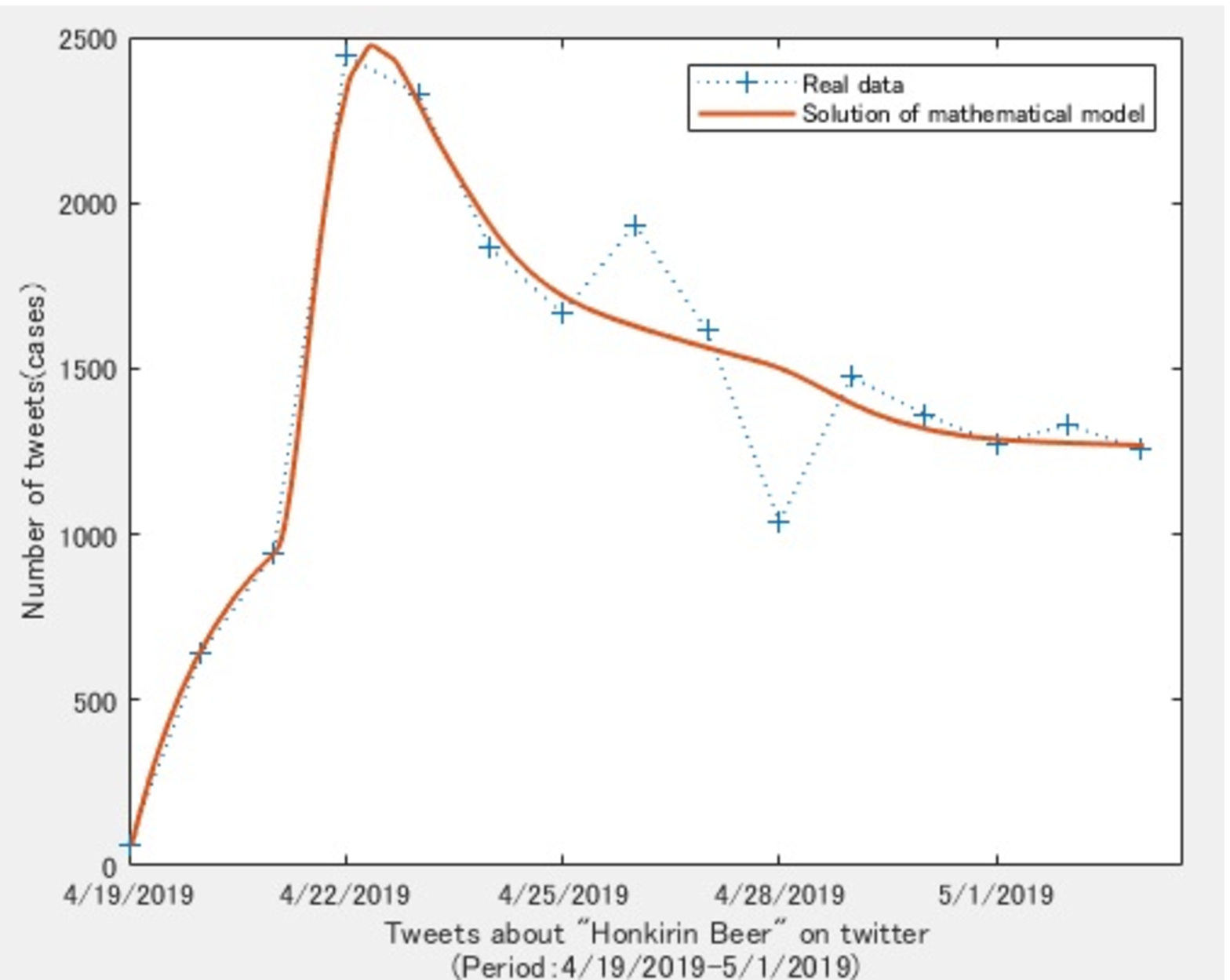}
  \caption{Fitting to Honkirin Beer data}
  \label{honkirin}
\end{figure}

The actual data in Figure \ref{honkirin} show a decline after the first peak slowly with one peak and one bottom on April 26 and 28 2019.
Similar to the other examples, the graph from the proposed model 
shows a high degree of fit to the first peak. 
In addition, we observe that 
it is able to fit the subsequent curves to the actual data except for 
two specific points. 
The high degree of accuracy in the fitness 
is evident from the large coefficient of determination, 
$\mbox{R}^2 = 0.9365$.
On the other hand, we observe that the graph from the proposed model isn't able to express well two specific points.
Future challenges include improvements to these points. 
\begin{figure}
\centering
  \includegraphics[width=50mm]{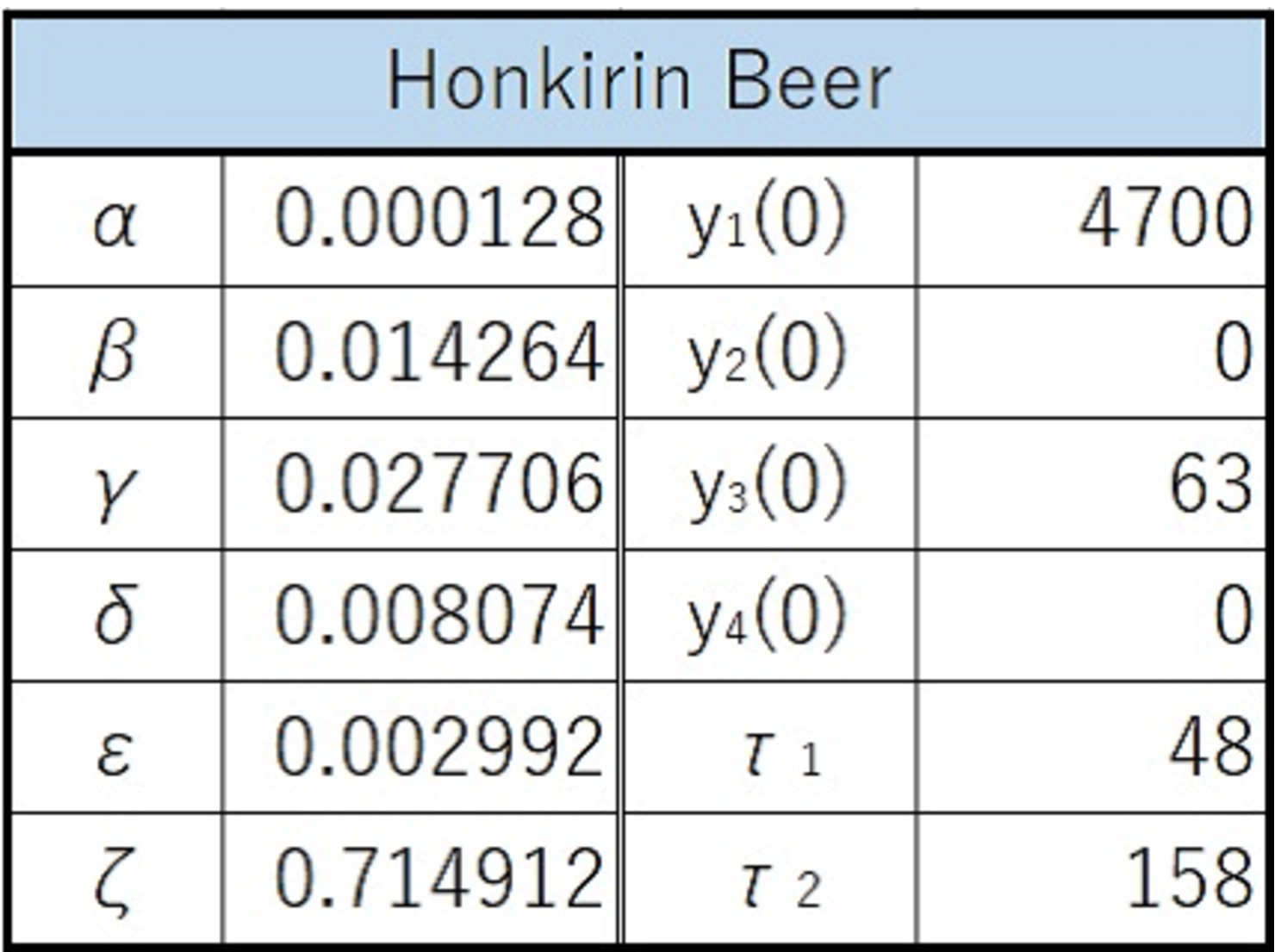}
  \caption{Various parameter values}
  \label{honkirin_para}
\end{figure}

\pagebreak

\subsubsection{Cup-noodle}

The first instant noodles were invented by Momofuku Ando, who later founded the well-known food company Nissin Food, in Japan in Osaka in 1958.
Currently, the instant noodles accomplish evolution of various form 
and are mainly sold as Cup-noodle at convenience stores and supermarkets.
In Japan, Cup-noodle is the most convenient quick meal and from 
the busy Japanese businessmen and women to the school children, it 
seems everyone enjoys them.
 
Here, 
we used Twitter data from 4/10/2019 to 5/3/2019 before and after 
the product presentation on 4/23/2019 to test the effectiveness of the cup-noodle presentation with respect to the fit of the proposed model to actual data. 

\begin{figure}[h]
\centering
  \includegraphics[width=80mm]{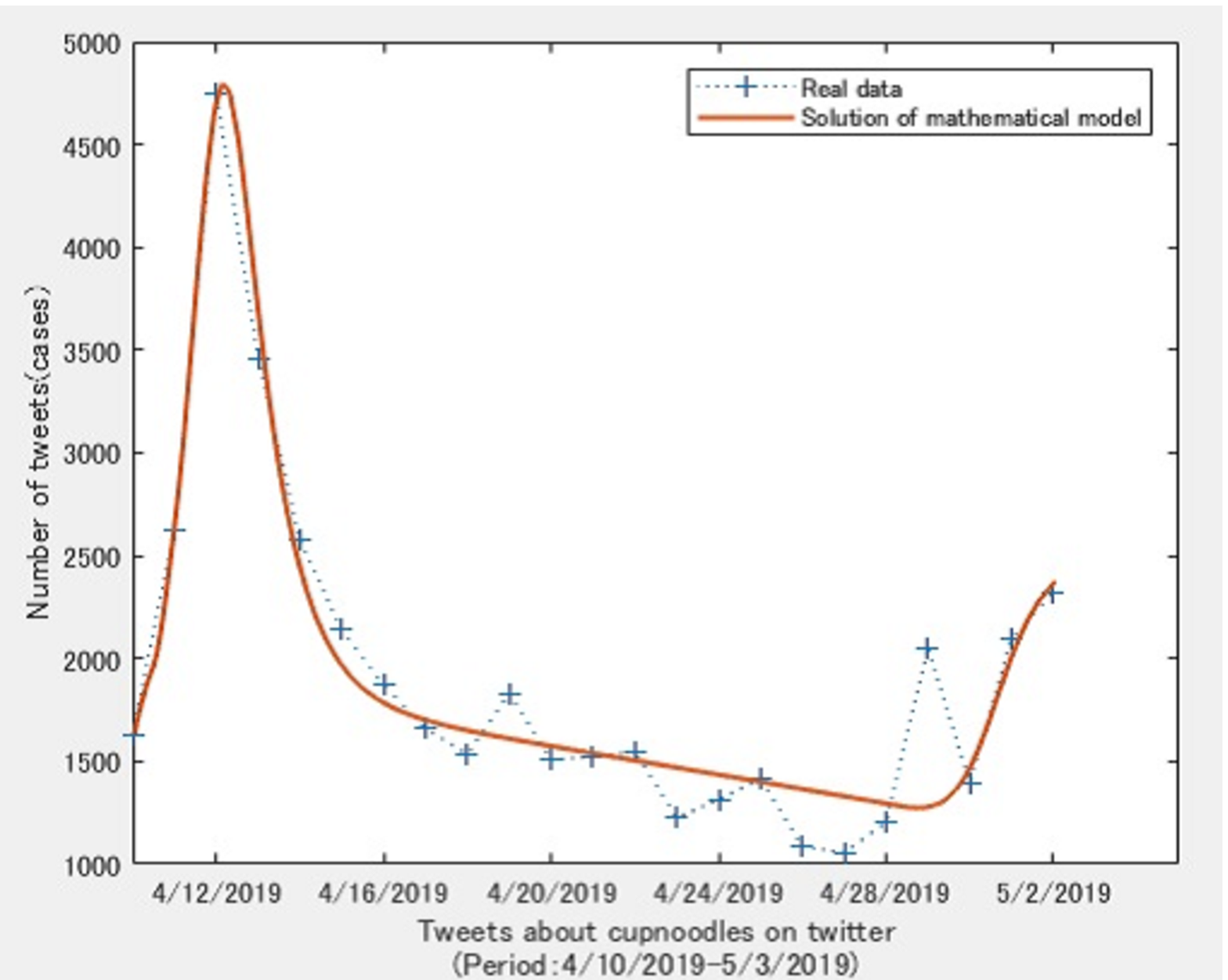}
 \caption{Fitting to Cup-noodle data}
  \label{cupnoodle}
\end{figure}

The actual data in Figure \ref{cupnoodle} show a decline after the first peak, and several vibrations and a sharp increase again towards the end day of this period.
Similar to the other examples, 
the graph from the proposed model shows a high degree of fit to the first peak. 
The high degree of accuracy in the fitness is evident from the large coefficient of determination, $\mbox{R}^2 = 0.9354$.
In particular, the graph of the proposed model expresses the rapid increase 
at the end day of this period.
These points are advantages of our proposed model.

\begin{figure}
\centering
  \includegraphics[width=50mm]{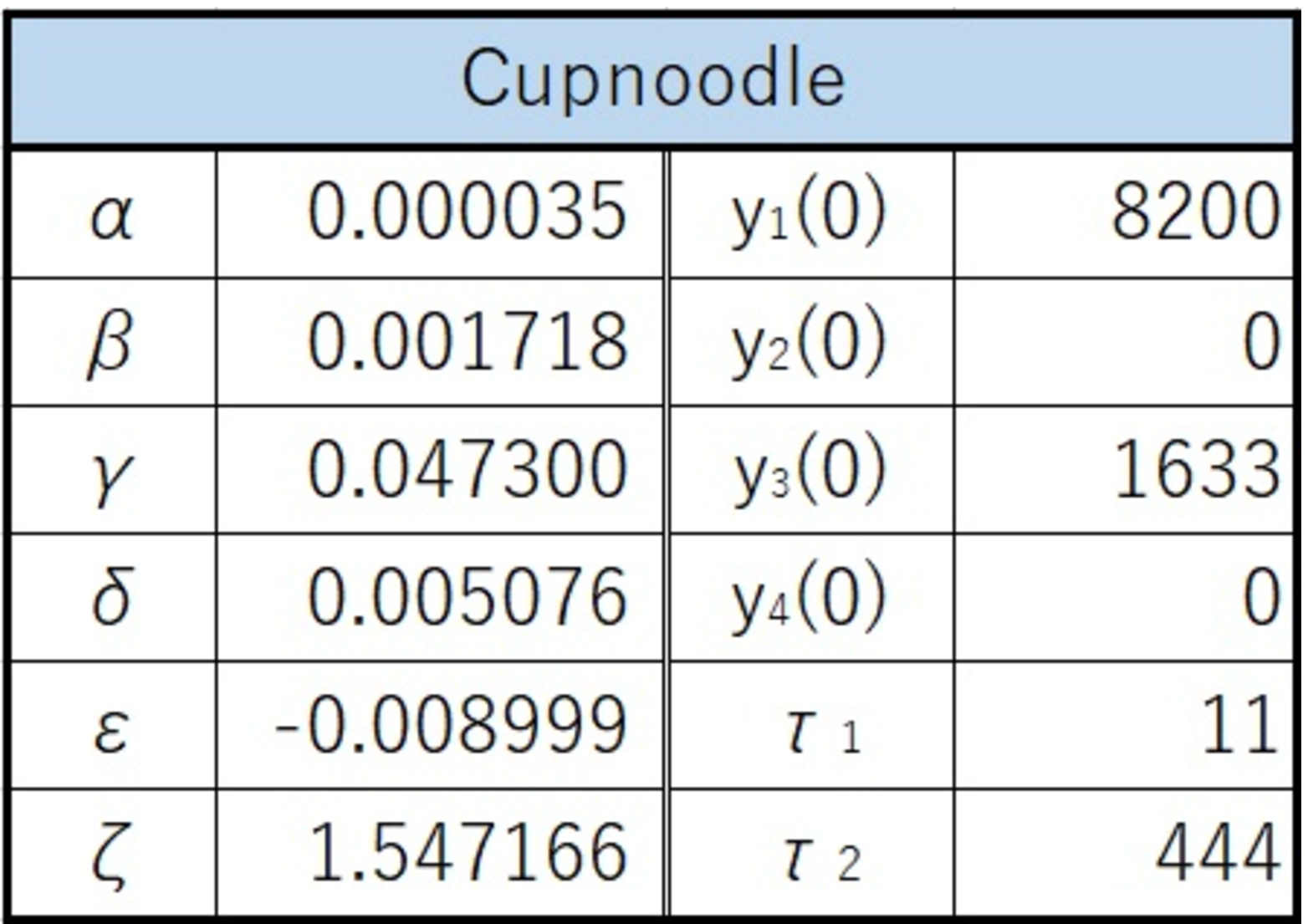}
  \caption{Various parameter values}
  \label{cupnoodle_para}
\end{figure}

\pagebreak

\subsubsection{``Reiwa (Beautiful Harmony)''}
In Japan, there are both the Christian era and the Japanese 
imperial era names are used, and when we have new emperor, 
new era starts.
There was a big event in Japanese royal family this year.
Current Emperor Akihito abdicated the throne on April 30 
and his son, Crown Prince Naruhito, enthroned the throne 
on May 1.
New Imperial era ``Reiwa (Beautiful Harmony)'' is taken from 
a verse in the oldest anthology of poems called Manyo-shu.
Here, we used Twitter data after the name of new Imperial era was announced on April 1 to test public interest of new Imperial era ``Reiwa'' for the fit of the proposed model to actual data. 

\begin{figure}[h]
\centering
  \includegraphics[width=80mm]{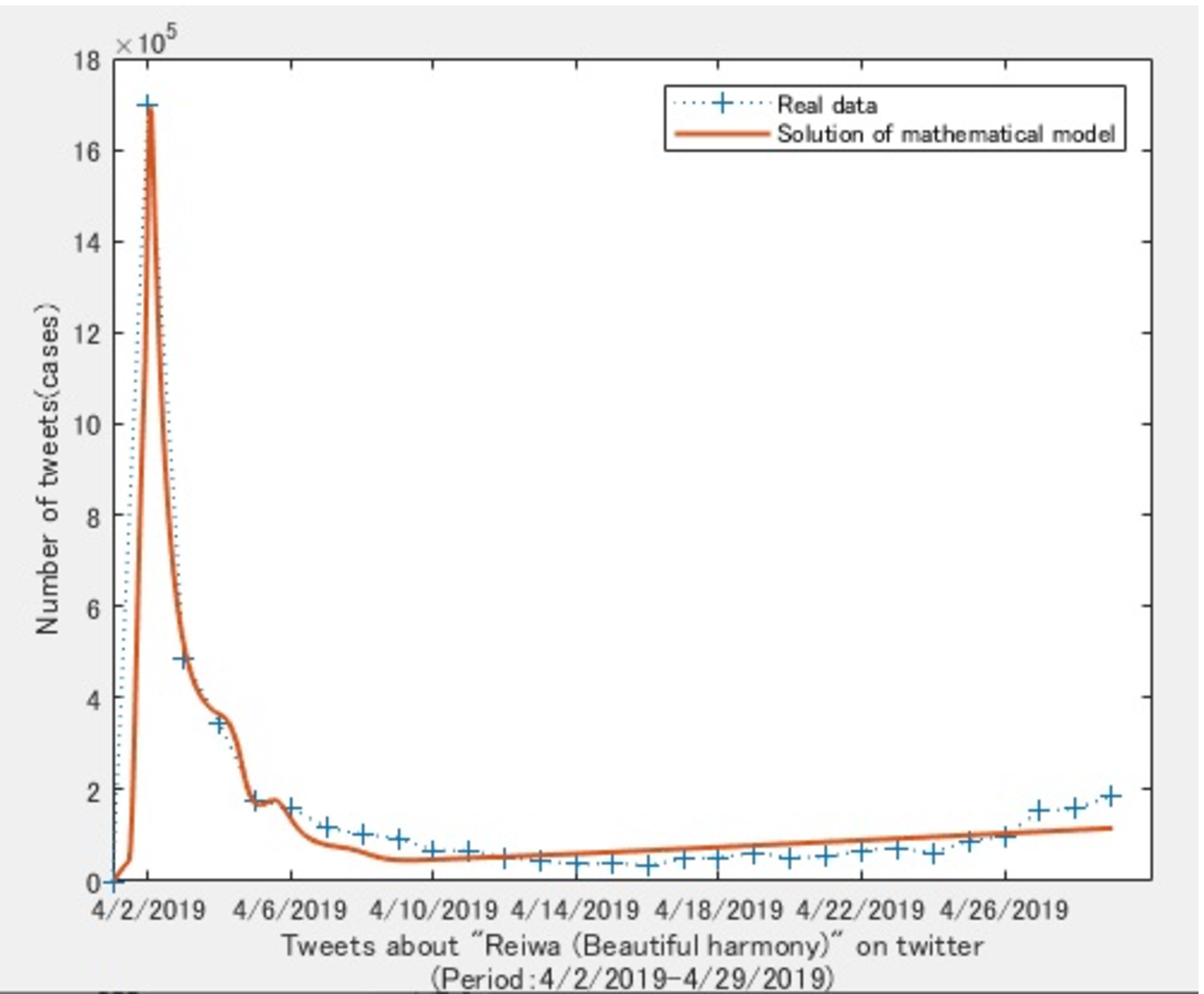}
\caption{Fitting to ``Reiwa (Beautiful Harmony)" data}
  \label{beautiful_harmony}
\end{figure}

The actual data in Figure \ref{beautiful_harmony} show a rapid, 
sharp spike followed by an easing, which is a characteristic pattern of transitory booms. 
The graph from the proposed model fits the actual data with a very high overall accuracy. 
The high degree of accuracy in the fitness is evident 
from the value of the coefficient of determination, $\mbox{R}^2 = 0.9910$.
In particular, the solution from the proposed model even expresses 
the small decline that occurs from 4/4/2019 to 4/6/2019. 
On the other hand, 
from the graph of the solution to the proposed model, 
we can see interests of Twitter users of new Imperial era ``Reiwa'' towards the date when the Era name changes from ``Heisei'' to ``Reiwa''.
Thus, we believe that an advantage of the proposed model is its ability to express vibrations by incorporating time delays and ``Sakura'' data.

\begin{figure}
\centering
  \includegraphics[width=50mm]{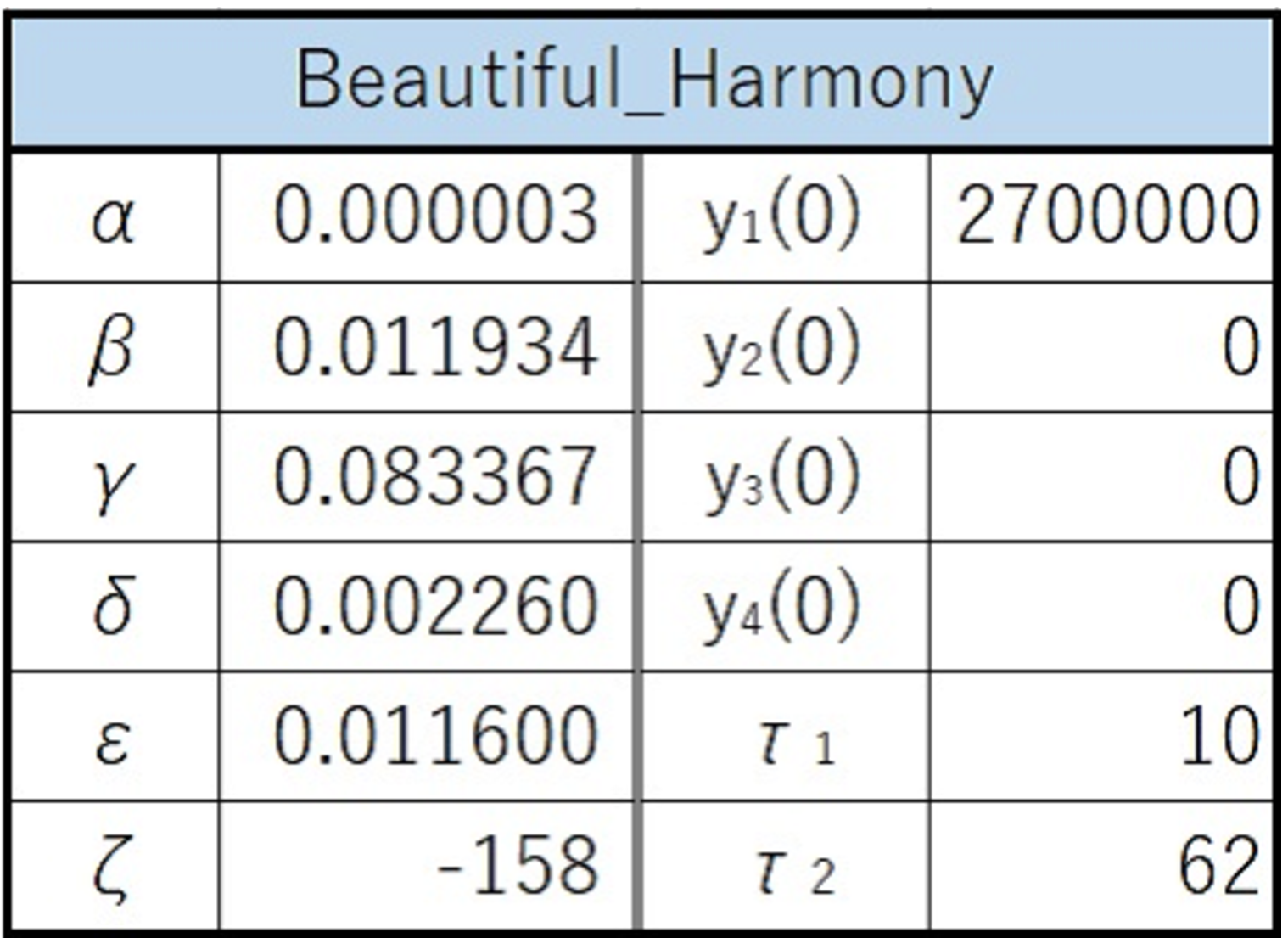}
  \caption{Various parameter values}
  \label{beautiful_harmony_para}
\end{figure}

\pagebreak

\section{Conclusion}
Based on our previous study, 
we examined the mathematical properties, 
especially the stability of the equilibrium 
for our proposed mathematical model in our previous study. 
By means of the results of the stability in this study, 
we also used actual data representing transient booms and resurgent booms, 
and conducted parameter estimation for the proposed model using Bayesian inference. 
In addition, we conducted a model fitting to actual data. 
By this study, we reconfirmed that 
we can express the resurgences or minute vibrations of actual data 
by means of our proposed model.

However, some issues remain. 
The first one is the validity of the proposed model. 
This study and our previous study only tested a total of 10 cases. 
In our future work, it will be necessary to take various examples of boom data, test the validity of our proposed model, and address the relationship between the actual data and parameters. 
Additionally, we would like to conduct an analysis using data not just for domestic booms but also global booms to test the validity of our proposed model. 
In addition, 
there is room for improvement in the parameter estimation method. 
In this study, 
we performed parameter estimation using Bayesian inference (the MCMC-MH method). 
In particular, 
there was a slight divergence in the second half of the graph and between the vibrations of the real data (actual vibration) and the vibrations of the mathematical model solution, as evinced by the fit to the resurgent boom data. 
This is considered to be a limitation in the model's reproducibility, 
but on the other hand, we will need to conduct testing in order to include improvements in the Bayesian inference method. 
In addition, there is room for improvement in how to determine 
$\tau_1$ and $\tau_2$, which represent time delays, and $\zeta$, which represents the presence of ``Sakura''. 
In our future work, we would like to search for parameters that have practical significance by incorporating findings from marketing research during the parameter estimation.
The expansion of the model is also an important topic. 
In particular, a mathematical model should be developed that incorporates probabilistic items to handle unexpected events. 
The ``Sakura'' parameter is a unique point of the proposed model, 
but realistically, a model is needed for scenarios other than that with a constant presence of ``Sakura''. 
We also must consider models with functions that have time-dependent parameters and time-dependent ``Sakura''. 

On the other hand, 
we must also advance mathematical analysis of areas such as the asymptotic behavior of mathematical models and their behavior around the points of equilibrium. 
In this study, we derived a sufficient condition with respect to the stability of the equilibrium for our proposed model, but it is certainly not necessary. 
In fact, the parameters to fit the actual data, ``Reiwa'' and ``Cup-noodle'' , didn't satisfy this sufficient condition.
However, fitting to the actual data is superior to several prior studies, and it is thought that our proposed model is superior for the means to express vibration and resurgence of the actual data during a certain period of time.
In our future challenges,  we improve the method of estimating parameters, by clarifying an association between mathematical properties and a Bayesian inference approach.

In recent years, 
the development of technology has made it relatively easy to obtain 
large data (both quantitatively and qualitatively), 
but a significant challenge has been how to use 
this large amount of data. 
We would like to continue research efforts in anticipation of 
the practical contribution of the mathematical model 
and parameter estimation method 
using a Bayesian inference approach in this study to corporate marketing strategies.

\vspace{5ex}

\noindent\textbf{Acknowledgments} \ 

The first author would like to acknowledge the supports from JSPS Grant-in-Aid for Scientific Research (C) 18K03439.
\vspace{7ex}

\bibliographystyle{unsrt}  


\end{document}